# ATMOSPHERIC SEASONALITY AS AN EXOPLANET BIOSIGNATURE

Stephanie L. Olson[1,2], Edward W. Schwieterman[1,2,3,4], Christopher T. Reinhard[1,5], Andy Ridgwell[1,2], Stephen R. Kane[1,2], Victoria S. Meadows[1,6], Timothy W. Lyons[1,2]

[1]*NASA Astrobiology Institute Alternative Earths and Virtual Planetary Laboratory Teams*
[2]*Department of Earth Science, University of California, Riverside, CA, USA*
[3]*NASA Postdoctoral Program, Universities Space Research Association, Columbia, MD, USA*
[4]*Blue Marble Space Institute of Science, Seattle, WA, USA*
[5]*School of Earth and Atmospheric Science, Georgia Institute of Technology, Atlanta, GA, USA*
[6]*Department of Astronomy, University of Washington, Seattle, WA, USA*

## ABSTRACT

Current investigations of exoplanet biosignatures have focused on static evidence of life, such as the presence of biogenic gases like $O_2$ or $CH_4$. However, the expected diversity of terrestrial planet atmospheres and the likelihood of both 'false positives' and 'false negatives' for conventional biosignatures motivate exploration of additional life detection strategies, including time-varying signals. Seasonal variation in atmospheric composition is a biologically modulated phenomenon on Earth that may occur elsewhere because it arises naturally from the interplay between the biosphere and time-variable insolation. The search for seasonality as a biosignature would avoid many assumptions about specific metabolisms and provide an opportunity to directly quantify biological fluxes—allowing us to characterize, rather than simply recognize, biospheres on exoplanets. Despite this potential, there have been no comprehensive studies of seasonality as an exoplanet biosignature. Here, we provide a foundation for further studies by reviewing both biological and abiological controls on the magnitude and detectability of seasonality of atmospheric $CO_2$, $CH_4$, $O_2$, and $O_3$ on Earth. We also consider an example of an inhabited world for which atmospheric seasonality may be the most notable expression of its biosphere. We show that life on a low $O_2$ planet like the weakly oxygenated mid-Proterozoic Earth could be fingerprinted by seasonal variation in $O_3$ as revealed in its UV Hartley-Huggins bands. This example highlights the need for UV capabilities in future direct-imaging telescope missions (e.g., LUVOIR/HabEx) and illustrates the diagnostic importance of studying temporal biosignatures for exoplanet life detection/characterization.

## 1. INTRODUCTION

The search for life beyond our solar system will focus on the identification of biosignature gases in exoplanet atmospheres or perhaps surface signatures of life (Des Marais et al. 2002; Schwieterman et al. 2018). The expectation that life should be recognizable via atmospheric chemistry on inhabited worlds is supported by the co-evolution of Earth's biosphere and atmosphere (reviewed by Olson et al. 2018), including the emergence of biogenic disequilibrium (Lederberg 1965; Lovelock 1965). Nonetheless, uniquely identifying life based on spectral snapshots is complicated by potential ambiguities, including false positives (abiotic processes that chemically mimic life) and false negatives (misleading non-detections). As an example, Earth's $O_2$-rich atmosphere is the most globally apparent expression of its photosynthetic biosphere (e.g., Lyons et al. 2014), and $O_2$ is thus the most widely referenced biosignature gas. However, high levels of $O_2$ do not always require biological processes (reviewed by Meadows 2017; Meadows et al. 2018): $CO_2$ photolysis or hydrogen escape may lead to detectable $O_2$ and/or $O_3$ (e.g., Domagal-Goldman et al. 2014; Luger & Barnes 2015). At the other extreme, $O_2$-producing life may evade detection. Oxygenic photosynthesis, for example, predated high levels of $O_2$ in Earth's atmosphere by more than two billion years if recent inferences of mid-Proterozoic (~1.8-0.8 Ga) $O_2$ levels are correct (Planavsky et al. 2014a, Planavsky et al., 2014b). Earth's oxygenation trajectory thus demonstrates that extensive biological production of $O_2$ in surface habitats need not result in detectable levels of $O_2$ in the atmosphere, and productive biospheres may be 'invisible' to conventional biosignature analysis (Gebuar et al. 2017; Reinhard et al. 2017). This potential for both false positives and negatives highlights the need to develop alternative biosignatures, including consideration of time-dependent signals.

Among a range of possibilities, seasonality in atmospheric composition is particularly promising. Atmospheric seasonality is biologically modulated on Earth and is likely to occur on other inhabited worlds regardless of specific metabolic substrates or products. It arises naturally on Earth, as it would elsewhere, from the interplay between the biosphere and axial tilt (obliquity). On Earth, for example, seasonal patterns in insolation shift the balance between photosynthesis:

$$CO_2 + H_2O \rightarrow CH_2O + O_2 \qquad [1]$$

and the reverse reaction, aerobic respiration:

$$CH_2O + O_2 \rightarrow CO_2 + H_2O, \qquad [2]$$

resulting in antithetic oscillations of atmospheric $CO_2$ and $O_2$ (Keeling et al. 1976). Meanwhile, net fluxes of $CH_4$ and other trace biological products evolve seasonally as well, responding to temperature-induced changes in biological rates, gas solubility, precipitation patterns, density stratification, and nutrient recycling (e.g., Khalil & Rasmussen 1983). Biologically modulated seasonality ultimately impacts nearly every constituent of Earth's atmosphere. The question is whether these time-variable signals would be detectable to a remote observer.

Despite detailed analysis of Earth's present-day atmospheric seasonality via ground-based and satellite observations, the discussion of seasonality as a biosignature has remained qualitative (Meadows 2006, 2008; Schwieterman 2018). We lack a comprehensive understanding of which spectral features are likely to be impacted by observable seasonality on inhabited worlds and how these impacts would be modulated by stellar, planetary, and biological circumstances. We also lack criteria for evaluating potential false positive seasonality scenarios, including insolation or temperature impacts on reaction rates or phase changes (e.g., Aharonson et al. 2009). We take a step here toward filling that gap by developing a conceptual framework that refines our expectations for, and interpretation of, temporal variability in atmospheric composition.

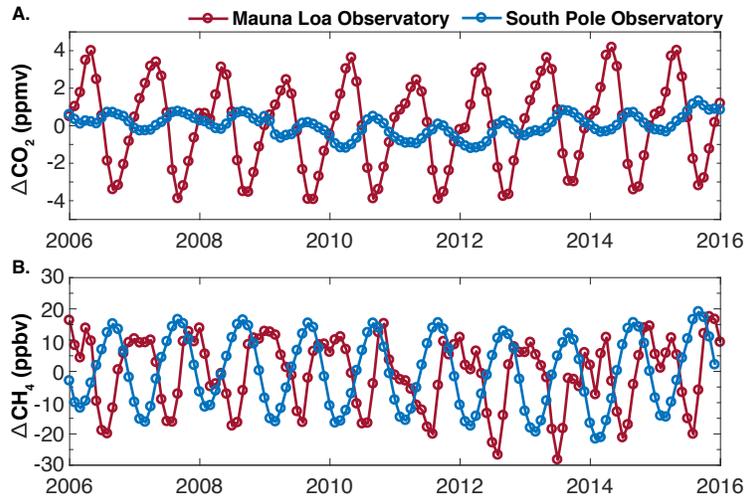

**FIGURE 1.** Detrended atmospheric $CO_2$ (**A**) and $CH_4$ (**B**) on monthly average from 2006-16 at Mauna Loa (19.5°N; red symbols) and South Pole (90°S; blue symbols) observatories. Data are from the National Oceanic and Atmospheric Administration (NOAA) Earth System Research Laboratory (ERSL) Global Monitoring Division (GMD; Dlugokencky et al. 2017a, b).

We begin by reviewing the biological, chemical, and physical mechanisms of $CO_2$, $CH_4$, and $O_2$ seasonality on Earth, including insight from its deep geologic past, and discuss how temporal variations in these gases may play out on terrestrial exoplanets. We include an examination of biotic and abiotic drivers of seasonal variations for these gases and how they may be discriminated. We end with a discussion of both the advantages and observational challenges facing seasonal biosignatures.

## 2. TARGET GASES

### 2.1 Carbon dioxide

Seasonal variability in atmospheric $CO_2$ is well-documented and mechanistically understood on the modern Earth (Keeling et al. 1976), and it is consequently the most frequently discussed temporal biosignature for Earth-like exoplanets (e.g., Schwieterman 2018). Carbon dioxide is also widely regarded as an important atmospheric component on habitable exoplanets owing to its likely role in climate regulation via temperature-dependent weathering feedbacks (Walker et al. 1981; Kopparapu et al. 2013). Assuming life elsewhere is also carbon-based, seasonal differences in biological C exchange with the environment may inevitably result in seasonality in atmospheric $CO_2$—independent of the specific metabolic substrates, pathways, and products involved. Therefore, $CO_2$ seasonality may be common to all inhabited worlds that experience seasonally variable insolation as the result of either orbital obliquity or eccentricity, although the apparent magnitude of biogenic seasonality will be modulated both by the land-ocean dichotomy of the planet and the viewing geometry of the observer (see **Section 5** below). From our perspective, the question is not whether biogenic $CO_2$ seasonality occurs elsewhere but whether it is detectable and if it can be discriminated from potential false positives.

Observing $CO_2$ seasonality will be a challenge for two reasons. First, $CO_2$ is extremely soluble in liquid water because it forms carbonic acid and other dissolved inorganic carbon species. Consequently, seasonal variations in $CO_2$ uptake and release by life in the ocean are chemically buffered, and these seasonal oscillations in the marine biosphere are ineffectively communicated to the atmosphere. The shifting balance of photosynthesis and respiration in land-based ecosystems, which are in more direct contact with the atmosphere, therefore dominate the seasonal $CO_2$ signal. Given this limitation, $CO_2$ seasonality would be an unsuitable biosignature for water worlds lacking substantial continental biospheres. Indeed, $CO_2$ seasonality is very weak in the ocean-dominated Southern Hemisphere today despite a strong signal in the more-continental Northern Hemisphere (**Fig. 1**). The second issue is that key $CO_2$ absorption features, such as the 4.3 and 15 μm bands, are effectively saturated at $CO_2$ levels that are low compared to the $CO_2$ levels that are likely on habitable zone (HZ) planets. The resulting loss of spectral sensitivity will further mute the spectral expression of $CO_2$ seasonality on many habitable planets, particularly at the outer edge of the HZ where atmospheric $CO_2$ might be very high (Kopparapu et al. 2013).

Attractive candidates for observable $CO_2$ seasonality would have terrestrial biospheres capable of producing a large magnitude seasonal signal but low overall $pCO_2$. An example of such a world may be Earth in the far geologic future following continued increases in solar luminosity and $CO_2$ drawdown via silicate weathering feedbacks (e.g., Lovelock & Whitfield 1982) or Earth-like exoplanets very near to the inner edge of their star's HZ.

Although seasonal $CO_2$ cycles may be vulnerable to non-detection (a false negative), large magnitude $CO_2$ seasonality is unlikely to arise from abiotic processes on habitable planets with liquid water, potentially making this biosignature robust against false positives. An exception to biological controls involves seasonal $CO_2$ ice sublimation and deposition, which produces a seasonal $CO_2$ cycle on Mars today (Wood & Paige 1992). However, significant levels of water vapor, which would suggest temperatures incompatible with seasonal $CO_2$ ice cycle, could preclude a Mars-like scenario. Seasonality in $CO_2$ in the presence of an ocean (Robinson et al. 2014) would be a powerful but difficult to detect indicator of life on habitable planets.

**2.2 Methane**
Seasonal oscillation in atmospheric $CH_4$ is also among the commonly discussed temporal biosignature (e.g., Schwieterman 2018). Although Earth's $CH_4$ is overwhelmingly biogenic, our seasonal $CH_4$ cycle is primarily photochemical. During the summer, warmer temperatures enhance evaporation. The resulting increase in tropospheric $H_2O$ promotes greater photochemical production of hydroxyl (OH) radicals, which accelerate $CH_4$ oxidation. Cold temperatures during the winter months have the opposite effect, muting $CH_4$ destruction. These seasonal changes produce the smooth $CH_4$ oscillation seen in the Southern Hemisphere (**Fig. 1B**). In contrast, the complex structure of Northern Hemisphere $CH_4$ seasonality arises from a biological $CH_4$ cycle on land, particularly at high latitudes, that is out of phase with the photochemical cycle (e.g., Khalil & Rasmussen 1983).

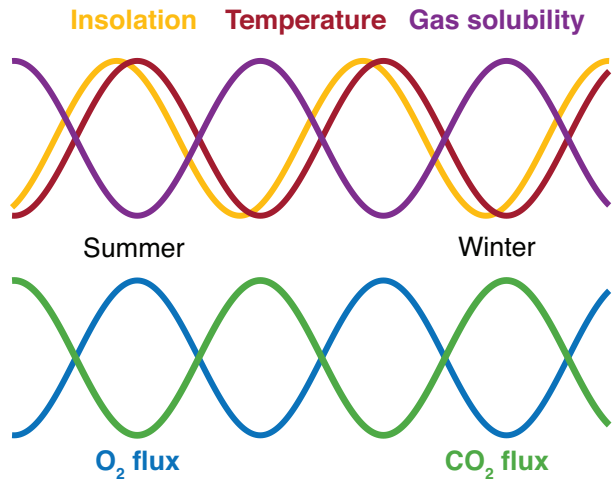

**FIGURE 2.** Schematic of the phase relationship between physical (top) and biological (bottom) factors impacting atmospheric seasonality in $O_2$ and $CO_2$. The temperature cycle is slightly out of phase with the insolation cycle because of the high heat capacity of water.

While Earth's seasonal $CH_4$ oscillation is largely an abiotic response to seasonality in tropospheric $H_2O$, it demonstrates the presence of a substantial surface liquid water reservoir—making it a habitability marker at the very least. A possible exception within our solar system involves seasonality in $CH_4$ evaporation, cloud formation, and precipitation as occurs on Titan (Aharonson et al. 2009; Coustenis et al. 2018), but this scenario is restricted to very cold bodies and is not relevant for HZ planets. These abiogenic signals complicate but do not preclude $CH_4$ seasonality as an independent biosignature. Indeed, Earth's seasonal recovery of $CH_4$ following elevated photochemical destruction requires a substantial flux from the planet's surface that, depending on broader planetary redox (Krissansen-Totton et al. 2018), would strongly imply a biological source of $CH_4$ to the atmosphere because it precludes stochastic delivery of exogenous $CH_4$ (Court & Sephton 2012) or episodic geologic inputs as occurs on Mars today (Mumma et al. 2009).

### 2.3 Oxygen & ozone

The cycles of $O_2$ and $CO_2$ are intimately linked through photosynthesis and aerobic respiration such that $O_2$ seasonality mirrors $CO_2$ seasonality (**Eq. 1-2**). We can predict from the above reactions that ~1 mole of $O_2$ increase balances ~1 mole of $CO_2$ drawdown, and vice versa, throughout the seasonal cycle. There are, however, important differences in the $O_2$ and $CO_2$ cycles that produce notable deviations from this idealized scenario—ultimately increasing the absolute (molar) magnitude of $O_2$ seasonality compared to that of $CO_2$.

Oxygen is significantly less soluble than $CO_2$, which increases the sensitivity of atmospheric $pO_2$ to seasonal changes in $O_2$ cycling at the Earth's surface—particularly within the surface ocean. The magnitude of $O_2$ seasonality is further amplified by the phase relationship between the net $O_2$ production/consumption cycle and the seasonal pattern of insolation, temperature, and thus gas solubility. Maximum $O_2$ production (and $CO_2$ drawdown) via photosynthesis occurs during the summer months when warm temperatures also minimize gas solubility (**Fig. 2**); the biological and physical components of seasonality therefore oppose each other for $CO_2$ but are additive for $O_2$. The magnitude of $O_2$ seasonality varies latitudinally (Keeling et al. 1998; Manning et al. 2003), but it is nearly 20x greater in absolute terms than the corresponding pattern for $CO_2$ in the present-day Southern Hemisphere (Keeling & Shertz 1992), albeit against a much higher background on the modern Earth. For these reasons, $O_2$ seasonality may, under some circumstances, have a greater potential to fingerprint the activities of aquatic biospheres, particularly those resembling Earth's early biosphere with its low levels of $O_2$.

Oxygen seasonality may also induce $O_3$ seasonality. Ozone is produced in the atmosphere photochemically from $O_2$ via (Chapman 1930):

$$O_2 + h\nu \rightarrow O + O \qquad [3]$$

$$O + O_2 + M \rightarrow O_3 + M, \qquad [4]$$

and $O_3$ is destroyed via:

$$O_3 + h\nu \rightarrow O_2 + O \qquad [5]$$

$$O_3 + O \rightarrow 2O_2. \qquad [6]$$

Ozone has a short atmospheric lifetime and its abundance is therefore strongly sensitive to both the UV environment and $O_2$ abundance (**Eq. 3-6**). Today's high levels of $O_2$ are not limiting for $O_3$ formation, and modern oscillations in $O_3$ are thus dominated by seasonally variable stratospheric transport (Weber et al. 2011; Butchart 2014). In a weakly oxygenated atmosphere, however, $O_3$ would be highly responsive to changes in $O_2$ (Kasting & Donahue 1980), possibly even the ppmv-level oscillation that we observe today (**Fig. 3A**). Thus, although $O_3$ is not a direct biological product, $O_3$ seasonality may be biologically modulated. Compared to the other seasonal cycles discussed here, $O_3$ seasonality may have a strong spectral impact, and we hypothesize that this signal may be detectable on weakly oxygenated worlds for which direct $O_2$ detection is unlikely.

## 3. CASE STUDY:
## SEASONAL EXPRESSION OF AN OTHERWISE CRYPTIC BIOSPHERE

As a proof of concept, we have quantified $O_3$ seasonality, chemically and spectrally, for a low-$O_2$ early Earth analog orbiting a Sun-like star. Our illustrative calculations consider an atmosphere with a long-term average $O_2$ level of 15 ppmv, compatible with geochemical proxies constraining $pO_2$ to $10^{-5}$ to $10^{-3}$ times the present atmospheric level (PAL) during the mid-Proterozoic, ~1.8-0.8 billion years ago (Pavlov & Kasting 2002; Planavsky et al. 2014b). We then impose modern magnitude $O_2$ oscillation in the ocean-dominated Southern Hemisphere around this baseline (Keeling & Shertz 1992; Keeling et al. 1998; Manning et al. 2003). This choice minimizes the influence of the terrestrial biosphere, which is appropriate because there were no land plants in the

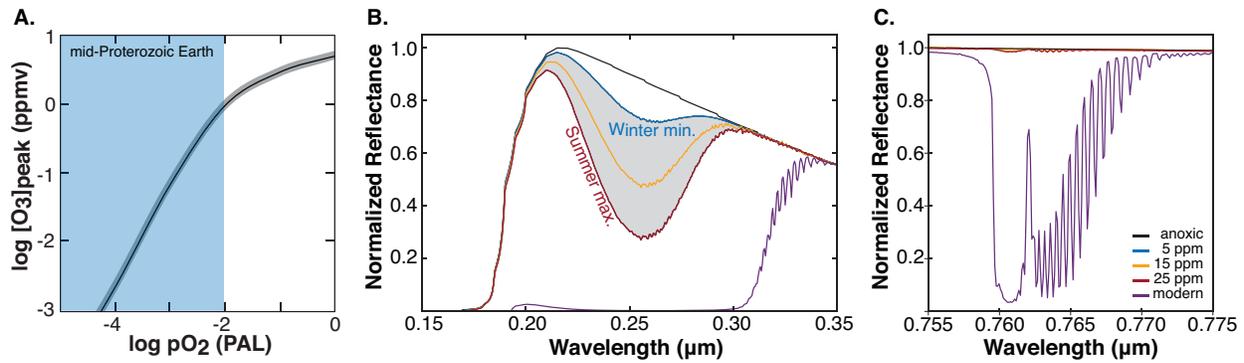

**FIGURE 3.** $O_3$ sensitivity to $pO_2$ (**A**) and synthetic spectra illustrating $pO_2$-dependent absorption by $O_3$ (**B**) and $O_2$ (**C**). Ozone is strongly sensitive to $pO_2$ (**A**; modified from Kasting & Donahue (1980)). Seasonal $O_2$ oscillation between 5-25 ppm imparts potentially observable $O_3$ seasonality in the Hartley-Huggins bands for a signal-to-noise ratio (S/N) of ~6 (**B**) but would be unobservable in the $O_2$-A band for S/N < 375 (**C**), assuming generous spectral resolving power of 500 at 0.25 μm and 7600 at 0.76 μm.

Proterozoic. The true magnitude of mid-Proterozoic seasonality is unknown, but the constancy of the carbon isotope record suggests that the productivity of the marine biosphere was not markedly different than today (Krissansen-Totton et al., 2015). Given that the magnitude of obliquity-induced insolation oscillation was also similar, modern magnitude seasonality in the Southern Hemisphere (~20 ppmv) is therefore a reasonable approximation for the mid-Proterozoic.

We calculated seasonality in $O_3$ abundance arising from this $O_2$ seasonality scenario using a 1D photochemical model (Kasting et al. 1979) that has a long legacy of modeling early Earth and Earth-like exoplanets (e.g., Arney et al. 2016, 2017). We then examined the spectral impacts of $O_2$ seasonality using synthetic reflectance spectra generated by SMART (Spectral Mapping Atmospheric Radiative Transfer; Meadows & Crisp 1996; Crisp 1997), focusing on the $O_2$-A band and the Hartley-Huggins $O_3$ bands. SMART has been validated by observations (Robinson et al. 2011) and was recently used to model the spectral appearance of early Earth analogs (Arney et al. 2016, 2017).

In our calculations, the seasonal $O_2$ maxima and minima result in distinctly different absorption signals for $O_3$ (**Fig. 3B**), but $O_2$ itself is masked in the spectra generated throughout the seasonal cycle (**Fig. 3C**). Consequently, biogenic $O_2$ seasonality may be inferred via $O_3$ seasonality, and biospheric $O_2$ fluxes could be quantified in the absence of detectable $O_2$.

Although photochemical scenarios for $O_3$ buildup are known and may display seasonality, seasonal recovery of $O_2$ requires an active source of $O_2$ commensurate with what is produced by the modern biosphere. This constraint rules out terminal atmospheric states where abiotic production and consumption fluxes of $O_2$ balance to zero, such as would be expected from extensive water loss in the geologic past. Additionally, it would be difficult to explain such a seasonal cycle in the absence of life on an Earth-like planet orbiting a Sun-like star because the abiotic production of $O_2$ is strongly disfavored around most FGK stars (see Harman et al. 2015, Domagal-Goldman et al. 2014). Confidence in a biological origin for $O_2$ and $O_3$ could be further enhanced by the detection of $H_2O$, strong Rayleigh scattering, and $N_2$-$N_2$ collisional absorption (Domagal-Goldman et al. 2014; Schwieterman et al. 2015) and by excluding significant CO, which would be expected to arise from $CO_2$ photolysis (Harman et al. 2015; Schwieterman et al. 2016). It is also important to note that inner-working angle constraints for future direct-imaging telescopes will favor the habitable zone planet-star combinations least conducive to hypothesized $O_2$ and $O_3$ false positives.

## 4. SEASONAL OSCILLATIONS: AN OPPORTUNITY AND OBSTACLE

Searching for seasonality as a biosignature may have several potential advantages and is highly complementary to conventional biosignature approaches. As discussed above, atmospheric seasonality may be a uniquely generic biosignature. This biosignature is likely to be particularly sensitive to photosynthetic biospheres, but its utility does not necessarily depend on the identity of the electron donor for photosynthesis (e.g., $H_2O$, $H_2S$, Fe(II))—and biospheres that do not harvest light energy may still produce atmospheric seasonality as they respond to oscillatory surface temperature. Seasonality may corroborate suggestions of life from isolated observations or mitigate potential false negatives through temporal variability in gases that are readily detectable in atmospheric spectra but are not uniquely biological in origin (e.g., $CO_2$, $CH_4$,). Even more intriguingly, seasonality may allow quantification of biospheric fluxes, providing an opportunity to characterize, rather than simply identify, exoplanet biospheres Like conventional biosignatures, seasonality will inevitably have both false positives and negatives, but it is nonetheless essential to develop biosignatures with differing biospheric blind spots to minimize the chance that we fail to recognize life elsewhere.

Despite the potential utility of seasonality in remote life detection, seasonality simultaneously poses observational challenges for remote life detection. An additional implication of our case study is that although $O_3$ may serve as a proxy for undetectable $O_2$ (e.g., Segura et al. 2003), $O_3$ and other short-lived gases might not be continuously detectable at every orbital phase in exoplanet atmospheres experiencing strong seasonality. Thus, it is possible that the spectral fingerprints of a biosphere may be missed in a single viewing. In the early Earth example presented here, intermittent detectability of $O_3$ would be particularly problematic because—in addition to low $O_2$—mid-Proterozoic Earth was also characterized by low $CH_4$ (Olson et al. 2016). Life was pervasive at the surface of the Earth at that time, but without access to spectroscopic data at UV wavelengths to support $O_3$ characterization (Reinhard et al. 2017), and potentially without repeated viewings, Earth may have appeared sterile to a remote observer expecting to detect $O_2$ or $CH_4$ on an inhabited planet. Consideration of seasonality is therefore important regardless of its utility as an independent biosignature because it may negatively impact the detectability of conventional biosignatures in isolated observations.

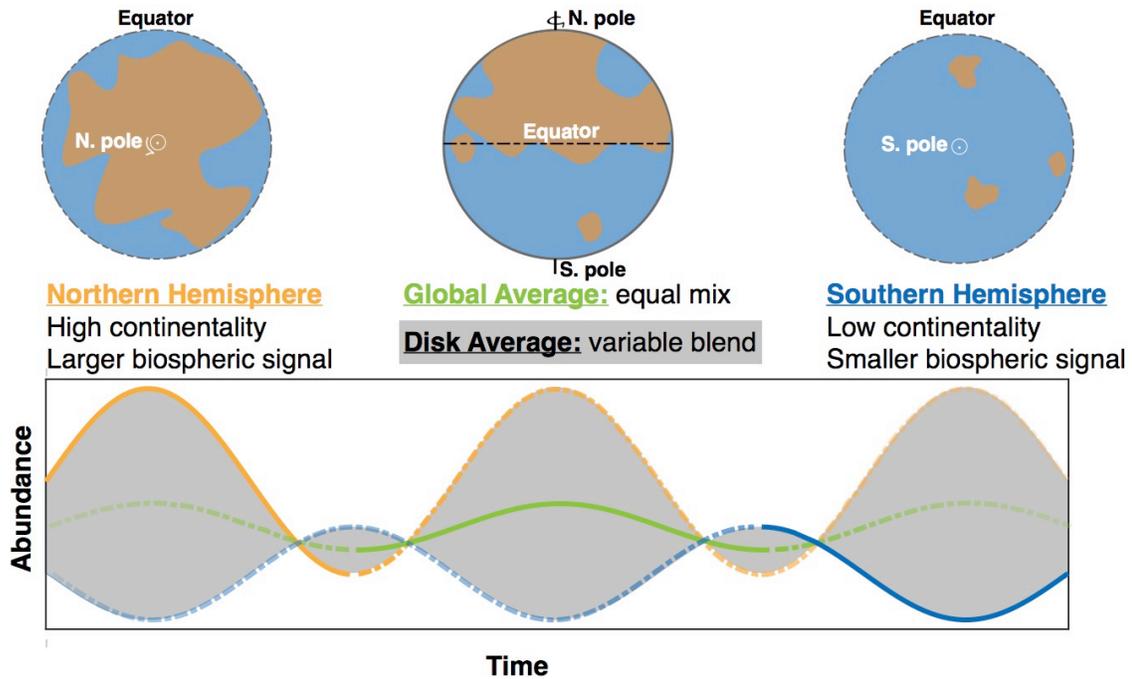

**FIGURE 4.** End-member viewing scenarios for a hypothetical exoplanet with asymmetric continentality (top), and the difference between hemispheric average, global average, and disk average (bottom). In the bottom panel, the Northern and Southern Hemispheric signals on this asymmetric planet are represented by the yellow and blue lines, respectively. The sum of these out-of-phase oscillations is the global average (green), but the observed signal in disk average will deviate from the global average within the shaded region depending on viewing geometry (**Fig. 5**).

## 5. OBSERVATIONAL CONSIDERATIONS

Characterization of seasonality demands a significant observational investment. Seasonal signals will be most readily accessible via direct imaging with future generations of large-aperture space-based telescopes (e.g., the LUVOIR or HabEx concepts; Dalcanton et al. 2015; Mennesson et al. 2016; Bolcar et al. 2017), as transit spectroscopy samples the same seasonal view, and thermal phase curves with JWST will likely lack the sensitivity to detect seasonal variations in gas abundances for terrestrial HZ planets (Meadows et al. 2018). Observing seasonality will also necessitate multiple viewings spanning months to years for Earth-like planets orbiting Sun-like stars, which will be among the most favorable targets for direct imaging studies because low angular separation will limit characterization of HZ planets around M stars.

In some cases, exoplanet viewing geometries may introduce challenges for exploiting seasonality as a biosignature. The perceived magnitude of seasonality will inevitably be less than the actual magnitude of seasonality at an exoplanet's surface owing to latitudinal averaging in disk-integrated spectra, including contributions from hemispheres experiencing opposite seasons (**Fig. 4**). Remote characterization of seasonality is thus limited by our ability to isolate opposing seasonal signals, as modulated by orbital obliquity and the inclination of the system with respect to the observer. The apparent seasonality may be further muted by planetary parameters, such as the distribution of continental area for some orbital configurations and viewing geometries. However, although asymmetry is important for generating seasonality on *global* average, such asymmetry is not strictly necessary for expressing seasonality in *disk* average (**Fig. 4**).

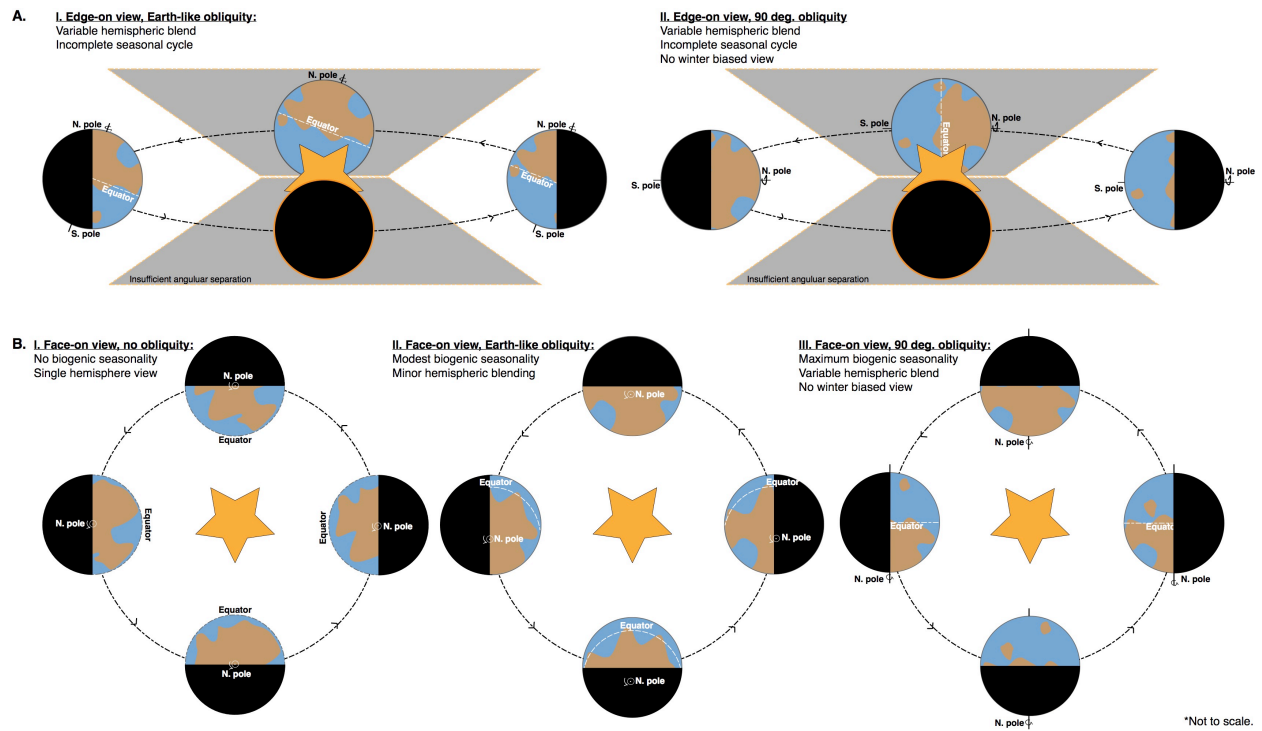

**FIGURE 5.** Modulation of seasonal signals by orbital obliquity for edge-on (**A**) and face-on (**B**) viewing scenarios. The magnitude of biogenic seasonality increases with obliquity, but high obliquity limits characterization of the winter hemisphere in both viewing configurations. For planets with Earth-like obliquity, face-on viewing will provide greater temporal coverage of the seasonal cycle and introduce less hemispheric blending compared to the edge-on scenario.

As orbital obliquity increases, the seasonal contrast—and thus the potential for large-magnitude biogenic seasonality—increases. However, while high obliquity maximizes the biogenic cycle, high obliquity also limits illumination of the winter hemisphere, potentially reducing the detectability of seasonal signals (**Fig. 5**). As the detectability of seasonality depends on both the magnitude of the biogenic signal and the extent to observation conditions mute that signal, the detectability of seasonality as a biosignature is likely optimized at intermediate obliquity that is sufficient to produce a large magnitude signal but not so large as to preclude viewing of the winter hemisphere. Future work leveraging 3D photochemical and spectral models will be required to offer robust quantitative predictions of these effects.

Global seasonality driven by eccentricity rather than hemispheric seasonality arising from obliquity introduces fewer geometric complications and may be more observable, so long as sufficient angular separation is maintained to observe summer at perihelion. The inclination of the system might also limit access to the full seasonal cycle, particularly for transiting exoplanets (**Fig. 5**), and may introduce challenges associated with variable hemispheric blending with evolving orbital phase. In these cases, seasonality may deviate from sinusoidal behavior. Characterization of seasonality will therefore be further optimized for a system with an inclination of ~0° with respect to the observer (a face-on scenario).

## 6. CONCLUSIONS & RECOMMENDATIONS

Atmospheric seasonality is a biologically modulated phenomenon on Earth, and biogenic seasonality may be common among inhabited worlds. As a biosignature, seasonality would be versatile with respect to metabolic sensitivity—and may even be an inevitable expression of a surface biosphere on planets that experience time-variable insolation. In addition to corroborating static evidence for life, seasonality may provide a tool to mitigate ambiguities (false positives and false negatives) and support characterization, rather than simple recognition, of an exoplanet biosphere. As an example, our calculations demonstrate that $O_2$ seasonality, as diagnosed via variations in the strength of the $O_3$ bands in the UV, may be the most notable spectral expression of life on weakly oxygenated planets, such as Earth's mid-Proterozoic biosphere. These exciting results suggest that quantification of $O_2$ fluxes may be possible even in the absence of detectable $O_2$ and strongly motivate the inclusion of UV (200-400 nm) capabilities in future direct imaging missions, such as LUVOIR or HabEx.

Ultimately, seasonality poses both an opportunity and a challenge for remote life detection because strong seasonality may render trace biological products only intermittently detectable. Continued investigation of biogenic and abiogenic seasonality for other gases, aerosols, or surface features—and their expression around different star types—will yield additional insight into the most likely spectral expressions of life in the Universe. Such studies will be essential to elucidate the value of repeated viewings spanning an exoplanet's full orbital period and inform the implementation of effective exoplanet viewing strategies for remote life detection.

**Acknowledgments**
The authors thank an anonymous reviewer and Jim Kasting for comments that substantially improved the manuscript. We acknowledge support from the NASA Astrobiology Institute under Cooperative Agreement Nos. NNA15BB03A and NNA13AA93A issued through the Science Mission Directorate. Additional support came from the NASA Astrobiology Institute's Director's Discretionary Fund and the NSF FESD program. EWS also acknowledges support from the NASA Postdoctoral Program, administered by the Universities Space Research Association.